\documentclass{article}

\usepackage[english]{babel}
\usepackage{microtype}
\usepackage{geometry}
\usepackage{setspace}
\usepackage{xcolor}
\usepackage{graphicx}
\usepackage{amssymb}
\usepackage{amsmath}
\usepackage{natbib}
\usepackage{float}
\usepackage{subcaption}
\usepackage{tabularx}
\usepackage[hidelinks]{hyperref}
\usepackage{authblk}

\title{Multiplex Hypergraph Modeling of Higher Order Structures in Psychometric Networks}
\author{
Francesca Possenti$^{1}$, Laura Girelli$^{3}$, Paolo Tieri$^{1,2}$, Manuela Petti$^{1}$\thanks{Correspondence: possenti.169567@studenti.uniroma1.it}\\
{\small $^{1}$DIAG, Sapienza University, Rome, Italy}\\ \small $^{2}$CNR Consiglio Nazionale delle Ricerche, IAC Istituto per le Applicazioni del Calcolo, Rome, Italy\\ \small $^{3}$University of Salerno, Italy
}

\date{} 

\begin{document}
\maketitle

\begin{abstract}
Psychiatric disorders have been traditionally conceptualized as latent conditions producing observable symptoms, but recent studies suggest that psychopathology may emerge from symptoms interactions. Psychometric networking model these relations focusing on pairwise associations but overlooks higher-order dependencies arising among groups of variables. These dependencies may reflect \textit{synergistic} mechanisms, where joint symptom configurations convey more information than pairwise relations, or \textit{redundancy}, where information overlaps.

We introduce an information-theoretic multiplex hypergraph framework to identify and compare higher-order interactions in eating disorders data, across diagnostic groups (e.g., anorexia nervosa). Higher-order structures are quantified using $\Omega$-information, a measure that captures the balance between redundancy and synergy. To address the combinatorial growth of candidate subsets, multiple testing and estimation instability, we propose a structured pipeline comprising: (i) targeted candidate selection based on dyadic network topology and theory-driven subscale information; (ii) a three-stage inferential procedure combining null-model testing with bootstrap robustness assessment; and (iii) the construction and analysis of diagnosis-layered, synergistic and redundant multiplex hypergraphs.

Results highlight how synergy captures the emergent, higher-order organization of diagnoses, revealing both a stable transdiagnostic core and diagnosis-specific ways in which these domains combine. By contrast, redundancy is confined to eating and body-image related content, marking reinforcement rather than broader symptom integration.

\end{abstract}

\section*{List of Abbreviations}

\begin{description}

\item[ANBP] Anorexia Nervosa, Binge--Purge subtype
\item[ANR] Anorexia Nervosa, Restricting subtype
\item[BED] Binge Eating Disorder
\item[BN] Bulimia Nervosa
\item[OSFED] Other Specified Feeding or Eating Disorder
\item[EDs] Eating Disorders
\item[EDI] Eating Disorder Inventory
\item[NSWD] Normalized Scale Weighted Degree

\item[ID] Interoceptive Deficits
\item[MF] Maturity Fears
\item[LSE] Low Self-Esteem
\item[IA] Interpersonal Alienation
\item[DT] Drive for Thinness
\item[B] Bulimia
\item[PA] Personal Alienation
\item[BD] Body Dissatisfaction
\item[A] Asceticism
\item[ED] Emotional Dysregulation
\item[II] Ineffectiveness
\item[P] Perfectionism
\item[EDI-3] Eating Disorder Inventory-3

\end{description}

\section*{Introduction}
Psychiatric disorders have long been viewed as latent disease entities that generate observable symptoms, a view formalized in diagnostic systems such as the Diagnostic and Statistical Manual of Mental Disorders (\textit{DSM-5-TR})~\citep{dsm5tr}. In contrast, accumulating theoretical and empirical work supports a systems perspective in which psychopathology emerges from interactions among symptoms ~\citep{borsboom2017network, mcnally2016brat}. This shift has motivated \emph{psychometric network science}, which models psychological constructs as networks of interacting components (e.g., symptoms or questionnaire items) rather than as reflective indicators of a small number of latent variables~\citep{borsboom2013net}. In these models, variables are represented as nodes and statistical relations as edges, enabling direct investigation of interdependencies, symptom-to-symptom pathways, and the organization of clinical phenomena~\citep{punzi2022review}.

Most applications in network psychometrics rely on graph-based models that encode \emph{pairwise} relations. A widely used approach is the partial correlation network, formally equivalent to a Gaussian graphical model, where an edge denotes non-zero conditional dependence between two variables controlling for the rest~\citep{epskamp2018estim}. 

Because psychometric data are often high-dimensional and exhibit substantial inter-item dependence, contemporary network estimation typically emphasizes sparse regularization and the assessment of parameter accuracy and stability \citep{friedman2008sparse,epskamp2018estim}.
Regularization methods such as the graphical lasso shrink weak connections to zero, yielding interpretable networks and reducing overfitting~\citep{friedman2008sparse}. Best practice further includes model selection procedures (e.g., EBIC or cross-validation) and robustness assessment via bootstrap resampling to quantify uncertainty in edge weights and centrality indices~\citep{epskamp2018estim}. These tools have supported substantive insights in psychopathology, including identification of central and bridge symptoms, community structure, and cross-construct linkages~\citep{punzi2022review}. 

Recent applied work also illustrates how network workflows can be used for questionnaire validation: for example, the analysis of the Eating Disorder Inventory 3~\citep{garner2004edi3} (\textit{EDI-3}) -- the same psychometric questionnaire analysed in this work -- combined redundancy screening (e.g., Unique Variable Analysis), regularized estimation, bootstrap aggregation, and community detection to recover meaningful item groupings while highlighting items with cross-loading network roles~\citep{christensen2020uva,punzi2023network}.

However, a central limitation of graph-based psychometric models is that they are restricted to dyadic interactions. Standard networks cannot represent dependencies that arise only at the level of three or more variables, even when such higher-order effects may reflect meaningful psychological mechanisms~\citep{benson2018simplicial, battiston2021natphys}. These include \emph{redundant} relations, where multiple variables share overlapping information, and \emph{synergistic} relations, where a configuration of variables jointly conveys information beyond what is present in any subset of pairwise links~\citep{marinazzo2024info}. 

To address this limitation, \citet{marinazzo2024info} introduced an information-theoretic framework that integrates multivariate dependency measures with hypergraph representations. In this approach, hyperedges connect sets of variables rather than pairs, allowing higher-order statistical structure to be directly encoded in the network architecture. A multivariate information metric is used to detect synergistic or redundant interactions among groups of variables -- the $\Omega$-information -- which are then mapped onto hyperedges that generalize conventional pairwise links. This formulation enables the simultaneous representation of pairwise and genuinely higher-order dependencies within a unified structure. The results presented by \citet{marinazzo2024info} demonstrate that such hypergraph constructions can reveal informative multivariate groupings, reduce confounding effects arising from overlapping indicators, and improve the recovery of meaningful modular organization compared with traditional pairwise network models.

At the same time, empirical hypergraph psychometrics raises practical challenges: the candidate search space grows combinatorially with the number of variables~\citep{choo2025hyper}, and stringent statistical control is required to reduce false discoveries and ensure robustness of detected higher-order effects~\citep{marinazzo2024info}. These considerations motivate structured pipelines that prioritize interpretable candidate sets, validate effects under appropriate null models, and assess stability via resampling. 

Against this methodological background, we build on established graph-based estimation practices~\citep{epskamp2018estim,friedman2008sparse} and recent information-theoretic hypergraph developments~\citep{marinazzo2024info} to investigate higher-order organization in psychometric data. Specifically, in this work, we analyze eating disorders (EDs) assessment data from the EDI-3 questionnaire to identify and compare redundant and synergistic interaction structures across different clinical groups. 

The EDI-3~\citep{garner2004edi3} is a standardized self-report questionnaire widely used to assess core symptoms and psychological features associated with eating disorders. It comprises 91 items (questions) rated on a $5$ points Likert scale, reflecting attitudes, behaviors, and emotional states. The instrument is organized into 12 primary subscales, covering both eating disorder–specific dimensions -- Drive for Thinness, Bulimia, and Body Dissatisfaction -- and broader psychological constructs, including low self-esteem, interpersonal difficulties, emotional dysregulation, interoceptive deficits, perfectionism, asceticism, and maturity fears. Together, these subscales provide a multidimensional profile of eating disorder psychopathology, capturing both symptomatic expressions and underlying psychological processes.

Eating disorders are severe psychiatric conditions defined in the \textit{DSM-5-TR} as persistent disturbances in eating or eating-related behavior that impair physical health or psychosocial functioning~\citep{dsm5tr}. Beyond disordered food intake, they involve broader disruptions in emotion regulation, cognition, and self-concept, and frequently co-occur with significant medical and psychiatric comorbidities. Their etiology is multifactorial, arising from interactions among biological vulnerabilities, psychological traits, and sociocultural factors, which contribute to heterogeneous clinical presentations and variable illness trajectories~\citep{feng2023eds, levinson2024guide, barakat2023risk}. 
EDs are associated with substantial psychological distress, medical complications, and elevated mortality risk~\citep{bhatta2020feed}. Moreover, from a systems perspective, eating-disorder symptomatology reflects interacting components -- such as restrictive behaviors, binge episodes, compensatory practices, body dissatisfaction, and affective dysregulation -- that may reinforce each other over time, contributing to symptom persistence and relapse~\citep{ortiz2026eds}.

Diagnostic classification has traditionally followed categorical criteria as defined in major nosological systems (e.g., DSM-5-TR), which distinguish between different EDs presentations based on behavioral patterns and associated psychological features. However, accumulating evidence suggests substantial overlap across diagnoses, motivating interest in dimensional and transdiagnostic approaches~\citep{eddy2008crossover,fairburn2003transdiagnostic}. 

Building on such assumptions, this study pursues a dual aim: to combine substantive psychometric questions with a methodological contribution. First, from a psychometric perspective, it seeks to investigate the role of individual EDI-3 items and subscales within each diagnostic group, to characterize synergy patterns at the subscale level, and to assess whether higher-order synergistic structures may carry diagnostic relevance across eating-disorder presentations. Second, from a methodological perspective, the study aims to advance psychometric network science by developing a reproducible hypergraph-based analysis pipeline and by examining higher-order interaction patterns across clinical groups within a shared measurement space. Methodologically, these objectives are addressed by modeling diagnosis-specific symptom interdependencies through a multiplex hypergraph framework, in which common questionnaire items are represented across separate diagnostic layers, allowing direct comparison of higher-order structures while preserving group-specific interaction profiles.

\section*{Results}
To characterize diagnosis-specific patterns and identify distinctive behavioral signatures of EDs, we analysed data from the open access project \textit{EDI-3, American Clinical Cases}~\citep{brookings2020edi3}, and we constructed two multiplex (multi-layer) hypergraphs~\citep{lotito2024multiplex}: one encoding synergistic interactions and the other capturing redundant interactions. The dataset consists of a large sample of 1206 female patients with diagnosis among the major Eating Disorders: Anorexia Nervosa - Binge Purging type (\textit{ANBP}, 259 observations); Anorexia Nervosa – Restrictive type (\textit{ANR}, 341 observations); Avoidant/Restrictive Food Intake Disorder (\textit{ARFID}, 2 observations): Binge Eating Disorder (\textit{BED}, 42 observations); Bulimia Nervosa (\textit{BN}, 404 observations); Other Specified Feeding or Eating Disorder (\textit{OSFED}, 155 observations); Unspecified Feeding or Eating Disorder (\textit{UFED}, 3 observations). 
Given the distribution of observations across diagnoses, we limited the analysis to four layers: ANBP, ANR, BN, and a merged BED/OSFED group combining patients diagnosed with BED or OSFED.

Eating disorders include several distinct but related clinical presentations. Anorexia Nervosa is marked by restrictive eating, significantly low body weight, fear of weight gain, and body-image disturbance, and includes a restrictive subtype (ANR) and a binge-eating/purging subtype (ANBP). Bulimia Nervosa involves recurrent binge eating followed by compensatory behaviors, usually without significant underweight. Binge-Eating Disorder is characterized by recurrent binge eating without regular compensation. Other clinically significant but atypical presentations are classified as OSFED/UFED.

Across diagnoses, symptom presentations frequently overlap and can shift over time, motivating transdiagnostic approaches that focus on shared maintaining mechanisms rather than rigid categories. Several psychological processes recur, including the \emph{overvaluation of weight and shape} (self-worth tied to body image and eating control), \emph{perfectionism} and \emph{cognitive inflexibility} (rigid rules and black-and-white thinking), and \emph{emotion regulation difficulties} (use of EDs behaviors to manage negative affect). Interpersonal sensitivity and social comparison further maintain symptoms, often alongside comorbid anxiety, depression, or obsessive-compulsive features. Together, these mechanisms suggest EDs reflect interacting cognitive--emotional--behavioral processes rather than isolated symptom clusters, supporting systems-oriented modeling approaches.

\subsection*{Modeling psychometric items interdependencies as multiplex hyperedges}
\begin{figure}[H]
    \centering
    \includegraphics[width=0.8\textwidth]{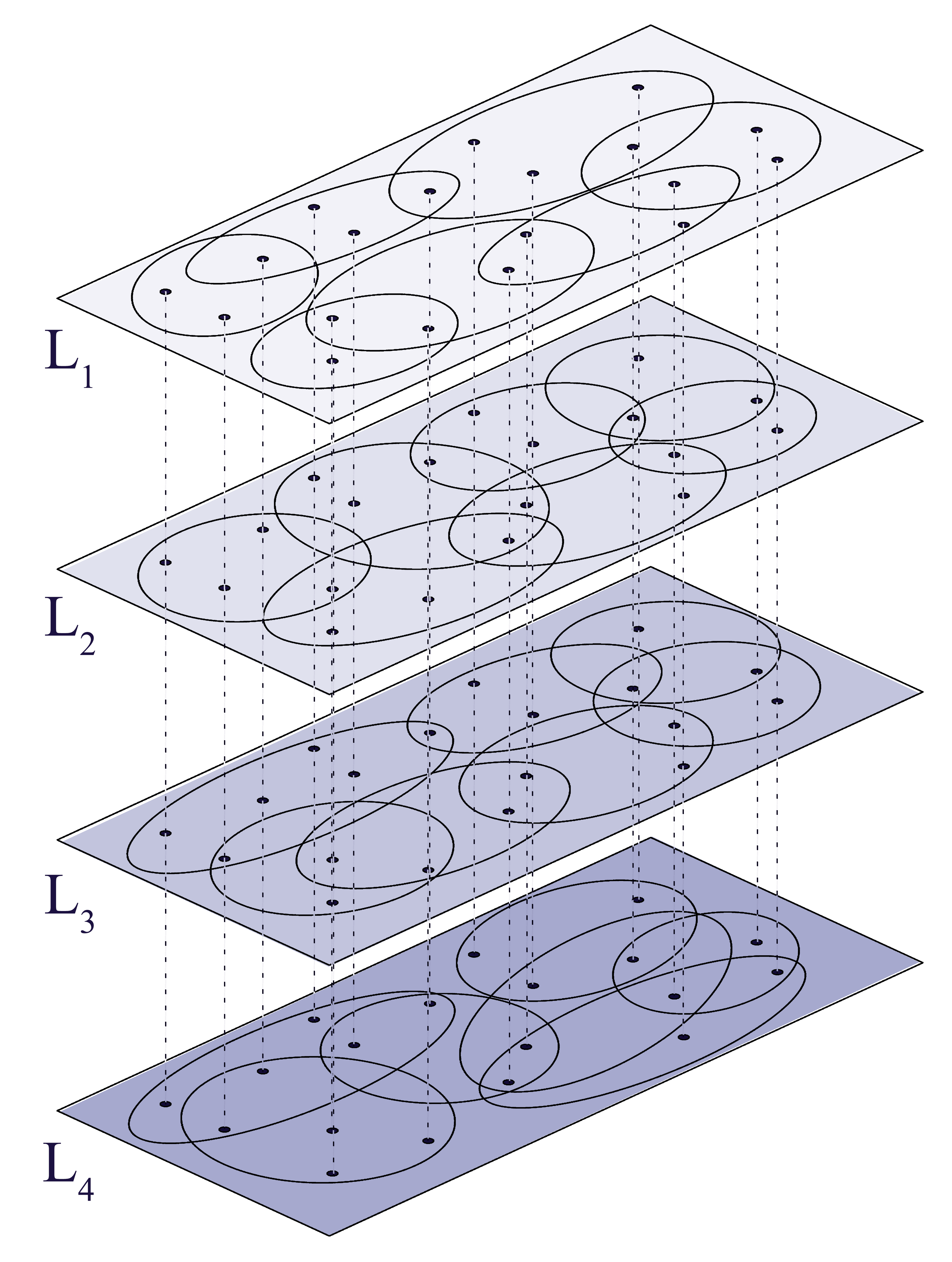} 
    \caption{\textbf{Multiplex hypergraph model of higher-order symptom interactions.} Schematic representation of the multiplex hypergraph framework used to model diagnosis-specific higher-order dependencies among EDI-3 items. Nodes represent questionnaire items and are shared across layers. Each layer corresponds to a distinct diagnostic group and contains its own set of weighted hyperedges, capturing higher-order interactions among items. Separate multiplex structures are constructed for synergistic and redundant interactions, allowing the decomposition of emergent versus overlapping informational patterns. This representation enables the comparison of item and scale-level organization across diagnoses within a unified higher-order network framework.}
    \label{fig:multiplex_hg}
\end{figure}

We constructed the synergistic and redundant multiplex hypergraphs~\citep{lotito2024multiplex}, where layers correspond to individual diagnostic categories and share a common set of nodes, while maintaining diagnosis-specific sets of hyperedges (Figure~\ref{fig:multiplex_hg}).

The 91 items of the EDI-3 questionnaire~\citep{garner2004edi3} were modeled as nodes, and higher-order statistical dependencies among items were represented as hyperedges. These higher-order interactions were estimated using the $\Omega$-information~\citep{rosas2019oinformation, marinazzo2024info}, a multivariate information-theoretic measure designed to quantify collective interdependencies beyond pairwise effects. In this framework, the $\Omega$-information also provides a natural weighting scheme for hyperedges, allowing each interaction to be characterized in terms of its synergistic or redundant contribution.

Among the 91 nodes considered in the system, not all participate actively across the multiplex layers. While the synergistic layers involve nearly the entire set of nodes -- with the exception of ANR, which does not include the \textit{non-scale item 77} -- only a relatively small subset contributes to the redundant multiplex structure. Specifically, the number of nodes active in redundancy is markedly lower across groups (ANBP: 22, ANR: 27, BED/OSFED: 23, BN: 15). 

\begin{figure}[H]
    \centering
    \includegraphics[width=0.8\textwidth]{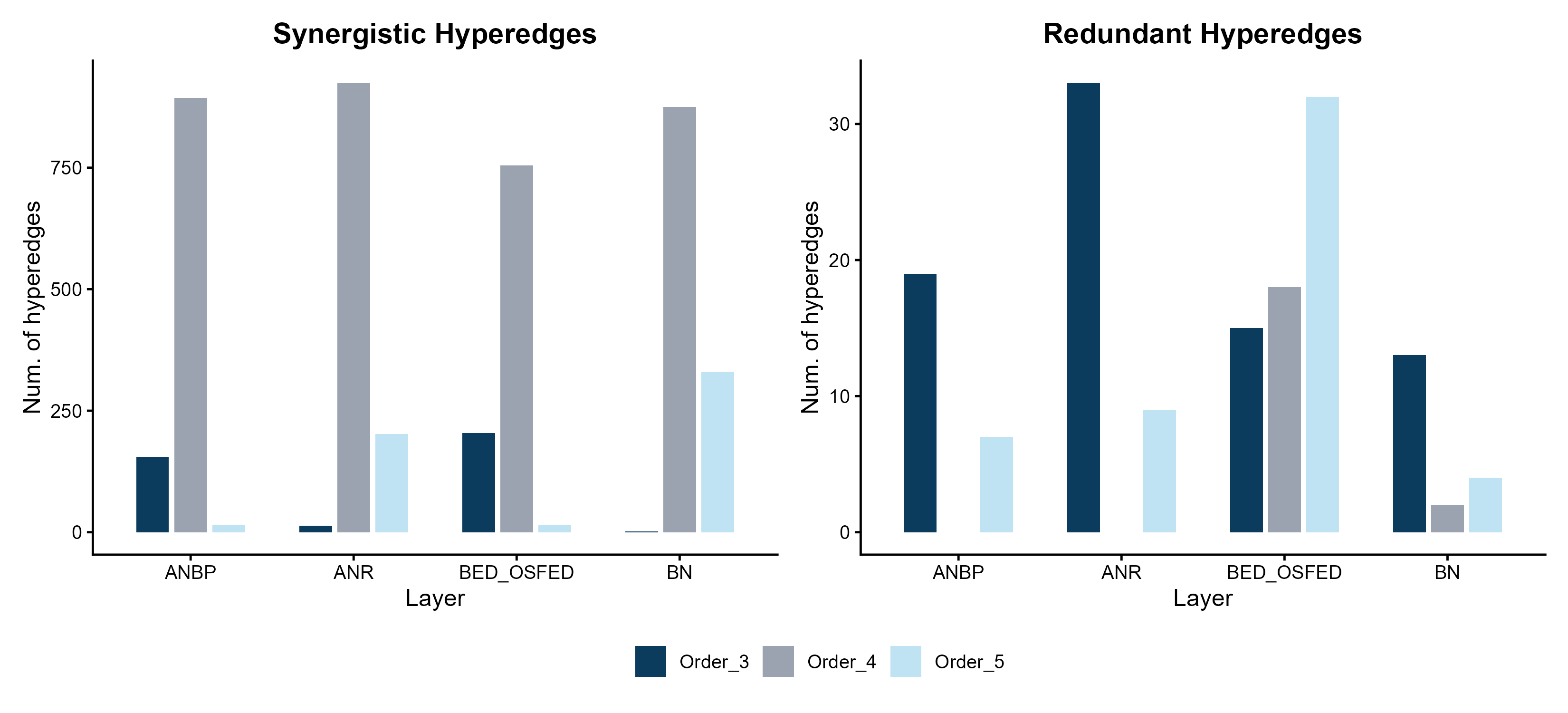} 
    \caption{\textbf{Distribution of synergistic and redundant hyperedges across interaction orders and diagnostic groups.} The number of hyperedges identified at interaction orders 3, 4, and 5 is shown for each diagnostic layer (ANBP, ANR, BED/OSFED, BN). The left panel reports synergistic hyperedges, while the right panel shows redundant hyperedges. Synergistic structures are markedly more numerous and are predominantly concentrated at order 4 across all groups, indicating that most synergistic dependencies arise from combinations of four variables. In contrast, redundant hyperedges are substantially less frequent and are mainly represented at order 3, reflecting overlapping informational contributions among smaller variable sets. An exception is observed in the BED\_OSFED layer, where redundancy is more prominent at order 5, suggesting the presence of higher-order overlapping dependencies in this group.}
    \label{fig:multiplex_hes}
\end{figure}

Another notable difference concerns the distribution of interaction orders. Synergistic structures are predominantly expressed at order 4, indicating that most synergistic dependencies emerge from combinations of four variables. In contrast, redundancy is mainly concentrated at order 3, suggesting that overlapping informational content is typically captured by lower-order interactions. An exception is observed in BED/OSFED, where order-5 interactions constitute the most represented redundant configurations (Figure~\ref{fig:multiplex_hes}).

We conducted two complementary analyses: a node-based analysis and a hyperedge-based analysis.

The node-based analysis aimed to identify the most relevant nodes -- corresponding to individual EDI-3 items -- and to rank the scales to which they belong. This approach allowed us to assess the contribution of specific items and subscales within each diagnostic layer. In particular, we examined both synergistic and redundant interaction profiles.

At the level of individual items, items that systematically participate in redundant higher-order interactions can be viewed as central nodes within the diagnostic architecture: their informational content is repeatedly shared across different item groupings, indicating that the construct they capture is deeply embedded in the disorder’s structure and may represent a core diagnostic feature.
In contrast, highly synergistic items were considered central relational elements, emerging as potential targets for a relation-based interpretation of symptom organization.

The hyperedge-based analysis focused instead on identifying diagnosis-specific patterns of scale co-involvement. These patterns were derived from the scale membership sets of hyperedges and classified into monoscale and multiscale configurations. To examine how subscales jointly convey diagnostically relevant information, we concentrated primarily on synergistic multiscale patterns, as these capture non-trivial cross-scale interactions. Monoscale patterns, by contrast, largely reflect expected within-scale associations and were therefore not the primary focus of the analysis.

Regarding redundancy, we specifically considered patterns composed of items belonging to both a single scale and multiple ones, spanning interaction orders from three to five. The former configurations are informative about internal scale coherence, while the latter -- identifying redundant structures -- were interpreted as reinforcing mechanisms, reflecting overlapping information that may stabilize and consolidate diagnostic features.

\subsection*{Most Influential Items and Scales in Synergistic and Redundant Relationships}
To quantify the involvement of individual items in higher-order interactions within each diagnostic layer, we computed the weighted degree of each node in both the synergy and redundancy multiplex hypergraphs. This measure captures the total strength of higher-order interactions associated with each item, providing an index of its informational relevance within a given diagnosis. To allow comparisons across layers, weighted degrees were normalized by the total interaction strength within each layer.

At the subscale level, we derived a summary measure by aggregating the normalized weighted degrees of items belonging to the same EDI-3 subscale. To control for differences in subscale size, we computed the average contribution per item and further normalized these values across subscales within each layer. The resulting metric, termed Normalized Scale Weighted Degree (NSWD), reflects the relative contribution of each subscale to the higher-order synergistic or redundant structure within each diagnostic group.

Synergistic interactions arise when groups of items jointly convey more information than the sum of their individual contributions or lower-order subsets. In this context, it reflects emergent symptom constellations in which combinations of responses capture intensified or more specific latent constructs, identifying integrative higher-order dependencies linking distinct psychological dimensions.

At the item level, synergy is consistently anchored in Interoceptive Deficits (ID), Maturity Fears (MF), and Drive for Thinness (DT). High-ranking items frequently reflect confusion about internal states, developmental insecurity, and weight-related concerns. These elements appear across all diagnoses, indicating that emotional dysregulation and core eating-related themes constitute transdiagnostic building blocks of higher-order organization. However, the relative prominence of additional content varies: ANBP integrates bulimic features with emotional and developmental vulnerability; ANR emphasizes restrictive and weight-focused concerns; BED/OSFED amplifies interoceptive disturbance; BN incorporates a stronger interpersonal component.
Notably, several items consistently rank highly across all layers -- items 8 (ID), 10 (LSE), 1 (DT), 6 (MF) --, despite belonging to different subscales (Table~\ref{tab:syn_items_all}).

\begin{figure}[H]
    \centering
    \includegraphics[width=0.8\textwidth]{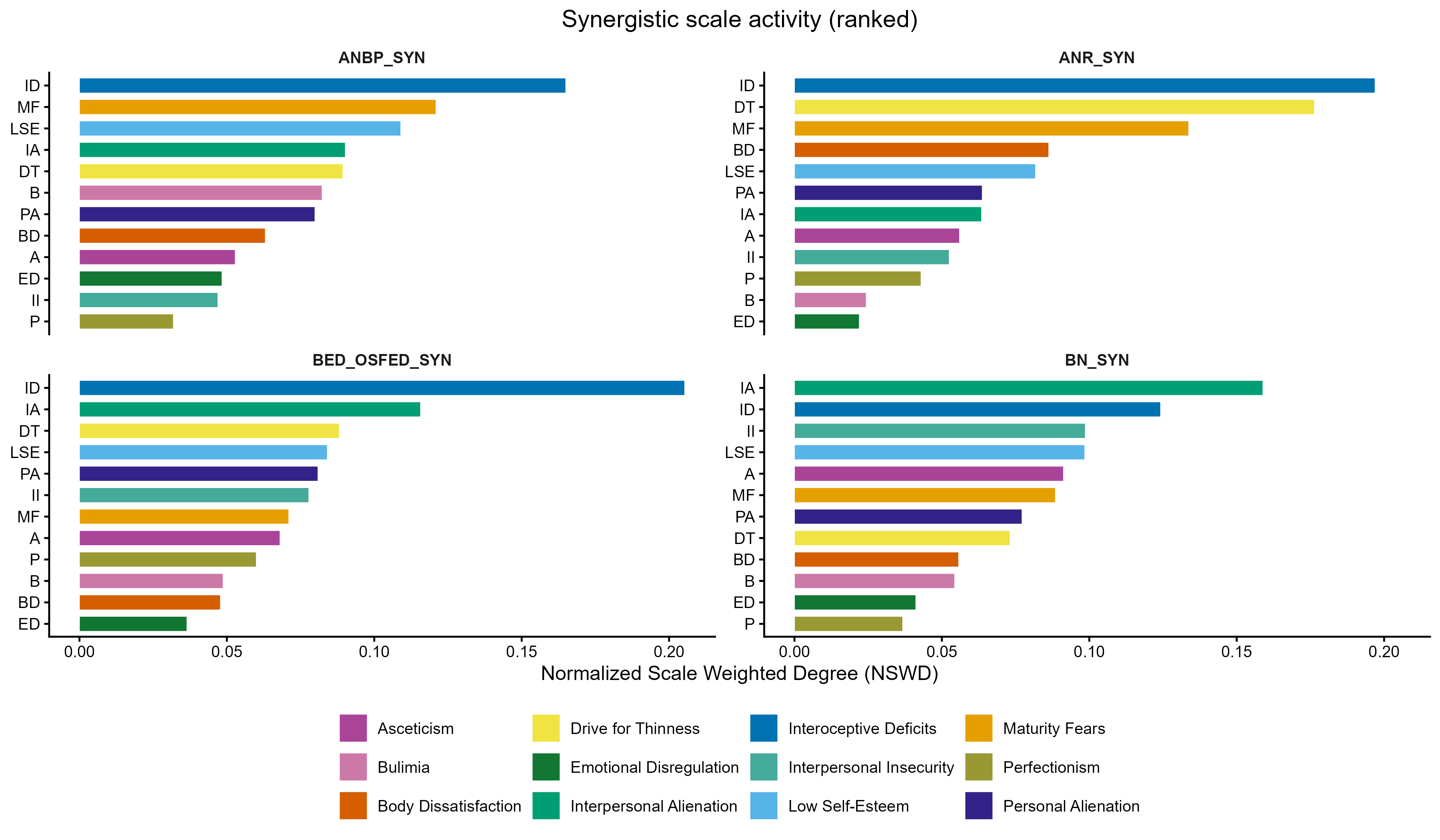} 
    \caption{\textbf{NSWD across diagnostic groups for synergistic interactions.} Bar plots display the normalized scale weighted degree for each EDI-3 subscale within the synergistic multiplex, separately for ANBP, ANR, BED/OSFED, and BN (four panels). Values represent the average per-item weighted degree within each subscale, further normalized within layer to allow comparison of relative higher-order involvement across scales. Bars are ranked within each panel. The figure highlights diagnosis-specific differences in the relative contribution of emotional, eating-related, and interpersonal dimensions to the higher-order synergistic structure.}
    \label{fig:synergy_scales}
\end{figure}

At the scale level, these differences are furtherly clarified: ANBP and ANR are primarily structured around ID and eating-related dimensions (DT, BD), with MF contributing substantially. BED/OSFED shows a marked dominance of ID, accompanied by Interpersonal Alienation (IA), suggesting a central role of interoceptive dysfunction reinforced by relational disconnection. In contrast, BN presents the most distributed configuration: although ID remains relevant, interpersonal scales (IA, II, PA) become comparatively dominant, indicating a reweighting of higher-order interactions toward relational functioning (Figure~\ref{fig:synergy_scales}).

Synergy reveals diagnosis-specific integration patterns that extend beyond single domains, linking emotional, eating-related, and interpersonal dimensions into distinct higher-order structures.

Redundant interactions occur when groups of items provide no additional information beyond their individual contributions. In the EDI-3, redundancy reflects tightly coupled symptom clusters characterized by overlapping variance and identify domains in which items are largely interchangeable because they capture the same underlying psychological dimension.

At the item level, redundancy is consistently dominated by eating and body related content. Across diagnoses, the most influential redundant items primarily belong to Body Dissatisfaction (BD), Drive for Thinness (DT) and Bulimia (B). In ANBP and BN, redundancy is almost entirely concentrated within BD, with top-ranked items referring to highly similar evaluations of specific body parts, indicating a strongly coherent body-image domain. In ANR, redundancy is shared mainly between DT and BD, reflecting substantial overlap between fear of weight gain, thinness preoccupation, and negative body evaluation. In BED/OSFED, redundant items extend beyond body dissatisfaction to include bulimic content, suggesting overlapping variance across binge-eating behaviors and body-image concerns (Table~\ref{tab:red_items_all}).

\begin{figure}[H]
    \centering
    \includegraphics[width=0.8\textwidth]{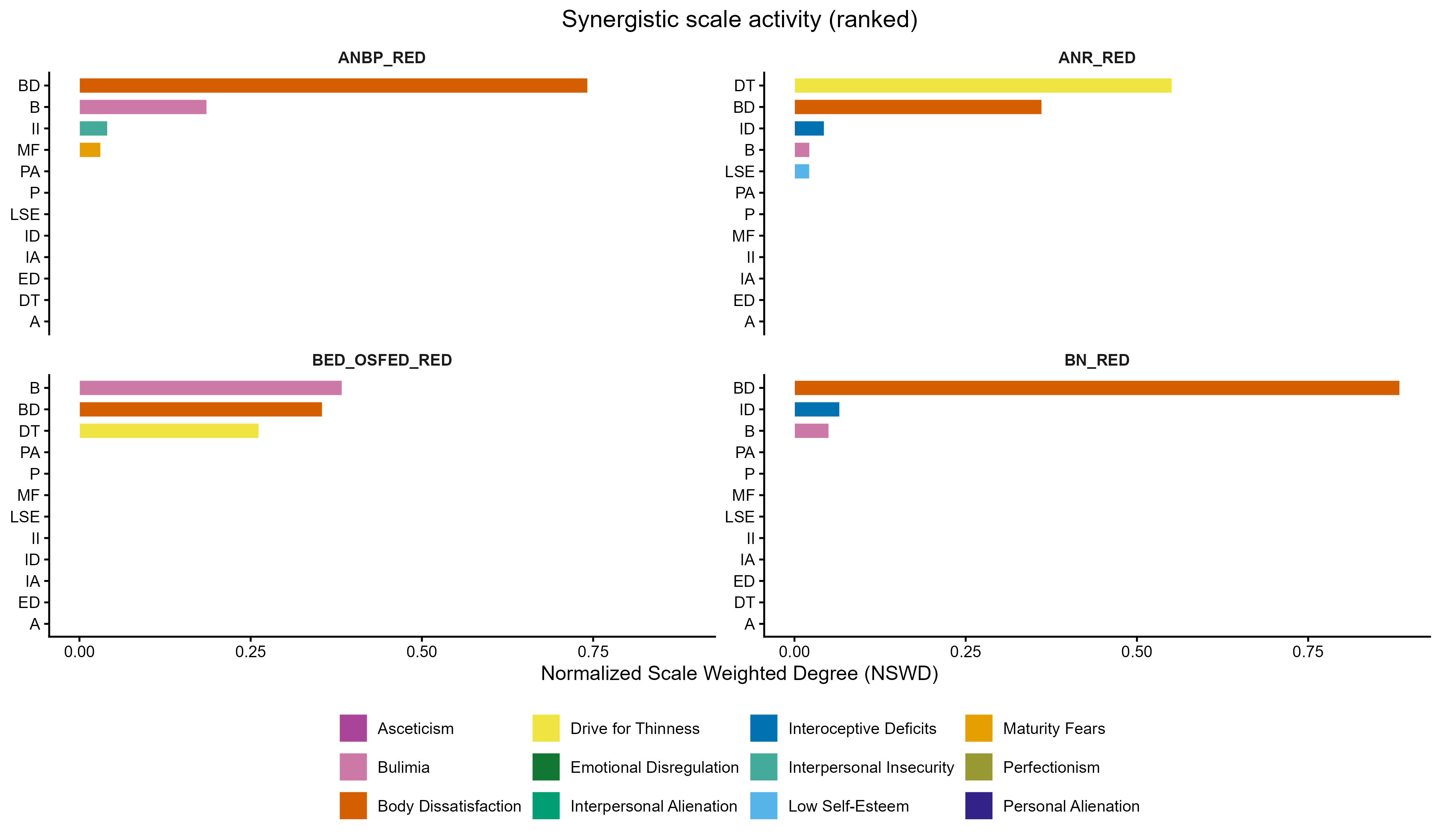} 
    \caption{\textbf{NSWD across diagnostic groups for redundant interactions.} Bar plots display the normalized scale weighted degree for each EDI-3 subscale within the redundant multiplex, separately for ANBP, ANR, BED/OSFED, and BN (four panels). Values represent the average per-item weighted degree within each subscale, further normalized within layer to allow comparison of relative higher-order involvement across scales. Bars are ranked within each panel. The figure highlights diagnosis-specific differences in the relative contribution of emotional, eating-related, and interpersonal dimensions to the higher-order redundant structure.}
    \label{fig:redundancy_scales}
\end{figure}

The same structure emerges at the subscale level: redundant weighted degree is highly concentrated within a small number of scales. ANBP and BN exhibit a pronounced dominance of BD, with only minor contributions from other dimensions. In both groups, B contributes marginally, while ID shows a limited additional role specifically within the BN group. ANR displays a slightly broader but still narrow structure, with redundancy largely confined to DT and BD. BED/OSFED presents a more balanced distribution across BD, B, and DT, yet redundancy remains restricted to eating-related domains (Figure~\ref{fig:redundancy_scales}).

Redundancy -- both at the item and scale levels -- is sharply constrained and domain-specific, capturing primarily overlapping information within closely related eating and body-image dimensions, in contrast to the more distributed and integrative patterns observed for synergy. This divergence underscores the fundamentally different roles of redundancy and synergy in structuring higher-order interactions within the EDI-3 context.

\subsection*{Scales Patterns in Synergistic and Redundant Hyperedges}
To systematically investigate interaction patterns at the subscale level, the analysis focused on higher-order dependencies among EDI-3 subscales. Rather than considering isolated items or dimensions, we examined patterns of joint involvement across psychological facets to identify which combinations of subscales systematically co-occur within higher-order structures, how these configurations vary across interaction orders and diagnostic layers, and whether they reflect diagnosis-specific or transdiagnostic organization. This approach provides a more integrated view of the multidimensional structure of the EDI-3 construct space, going beyond individual item-level effects.

\begin{figure}[H]
    \centering
    \includegraphics[width=0.8\textwidth]{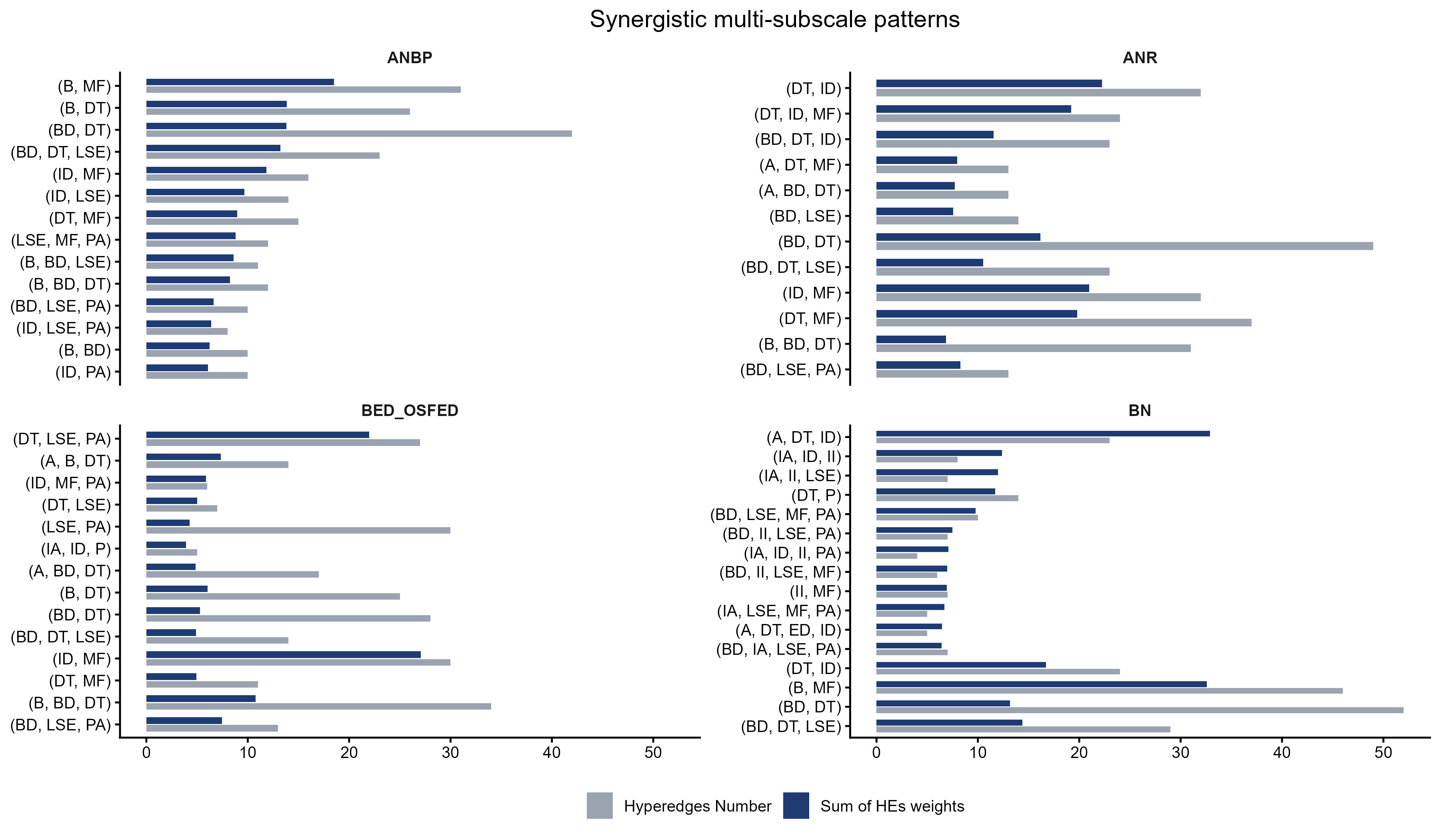} 
    \caption{\textbf{Synergistic multi-subscale interaction patterns across diagnostic layers.} Four panels display the most prominent multi-subscale synergistic configurations for ANBP, ANR, BED/OSFED, and BN. For each pattern, the total number of validated hyperedges, as \textit{Hyperedges Number} is shown in grey, while the cumulative interaction weight is shown in blue as \textit{Sum of HEs weights}. Patterns are ordered within each layer by the absolute magnitude of cumulative weight. Synergistic interactions are predominantly concentrated at intermediate-to-high hyperedge orders and vary systematically in composition across diagnoses, highlighting differences in higher-order integration among eating-related, interoceptive, self-evaluative, and interpersonal dimensions.}
    \label{fig:syn_pat_4panels}
\end{figure}

Regarding \textit{synergistic multi-scale patterns}, we examined higher-order configurations across diagnostic layers, focusing on patterns involving multiple EDI-3 subscales (Figure~\ref{fig:syn_pat_4panels}).

In \textit{ANBP}, synergy is structured primarily at order 4, with recurrent combinations linking eating-related subscales (B, BD, DT) to interoceptive, self-evaluation and developmental dimensions (ID, LSE, MF). Frequent and high-weight patterns such as (B, MF), (B, DT), and (BD, DT), along with higher-order configurations including LSE and ID, indicate a central role of B in this pathology and a generally organized interaction structure between core eating symptoms and affective or self-related components. The repeated involvement of MF and LSE suggests that developmental concerns and self-worth processes participate systematically in these higher-order configurations, while patterns including PA further point to an integration with interpersonal or identity-related dimensions. Overall, synergy in ANBP appears relatively coherent and anchored around bulimic symptomatology interacting with broader emotional and self-evaluative mechanisms.

\textit{ANR} shows a similarly structured profile, but centred on DT, ID, and MF. Recurrent pairs and triplets -- in particular (DT, ID), (ID, MF), (DT, MF), and the higher-order configuration (DT, ID, MF) -- suggest that synergy is driven by the joint engagement of restrictive eating concerns and internal emotional processing. Compared with ANBP, the structure appears more strongly anchored in restrictive pathology rather than bulimic symptomatology. The occasional involvement of LSE and A indicates that cognitive and behavioural features related to self-discipline and self-perception participate in these configurations, but they play a secondary role relative to the DT–ID–MF triad. Overall, the synergistic organization in ANR appears more internally focused, reflecting interactions between restriction, emotional awareness, and developmental concerns.

In \textit{BED/OSFED}, the synergistic structure becomes more heterogeneous. Although patterns involving ID and MF remain prominent, combinations span a broader mixture of eating, self-evaluative, and interpersonal dimensions. Configurations such as (DT, LSE, PA) and (BD, LSE, PA) illustrate the involvement of interpersonal and self-worth components alongside eating-related variables, while the presence of both bulimic (B) and body dissatisfaction (BD) indicators suggests multiple symptomatic pathways. Compared with the anorexia subtypes, the structure appears less centred on a single dominant configuration, instead reflecting a more distributed organization across psychological domains.

\textit{BN} exhibits the most integrated and complex structure. Many configurations involve higher interaction orders (frequently {4,5}) and combine eating-related, interpersonal, and emotional subscales. Patterns linking ID, DT, IA, II, LSE, and MF indicate a densely interconnected set of higher-order dependencies spanning emotional awareness, interpersonal functioning, and self-evaluation. Several configurations also integrate behavioural and affective components simultaneously, suggesting that bulimic symptomatology is embedded within a broader network of psychological processes rather than forming isolated interactions. Overall, BN displays the richest synergistic organization, characterized by extensive cross-domain integration and higher-order interaction complexity relative to the other clinical groups.

\begin{figure}[H]
    \centering
    \includegraphics[width=0.8\textwidth]{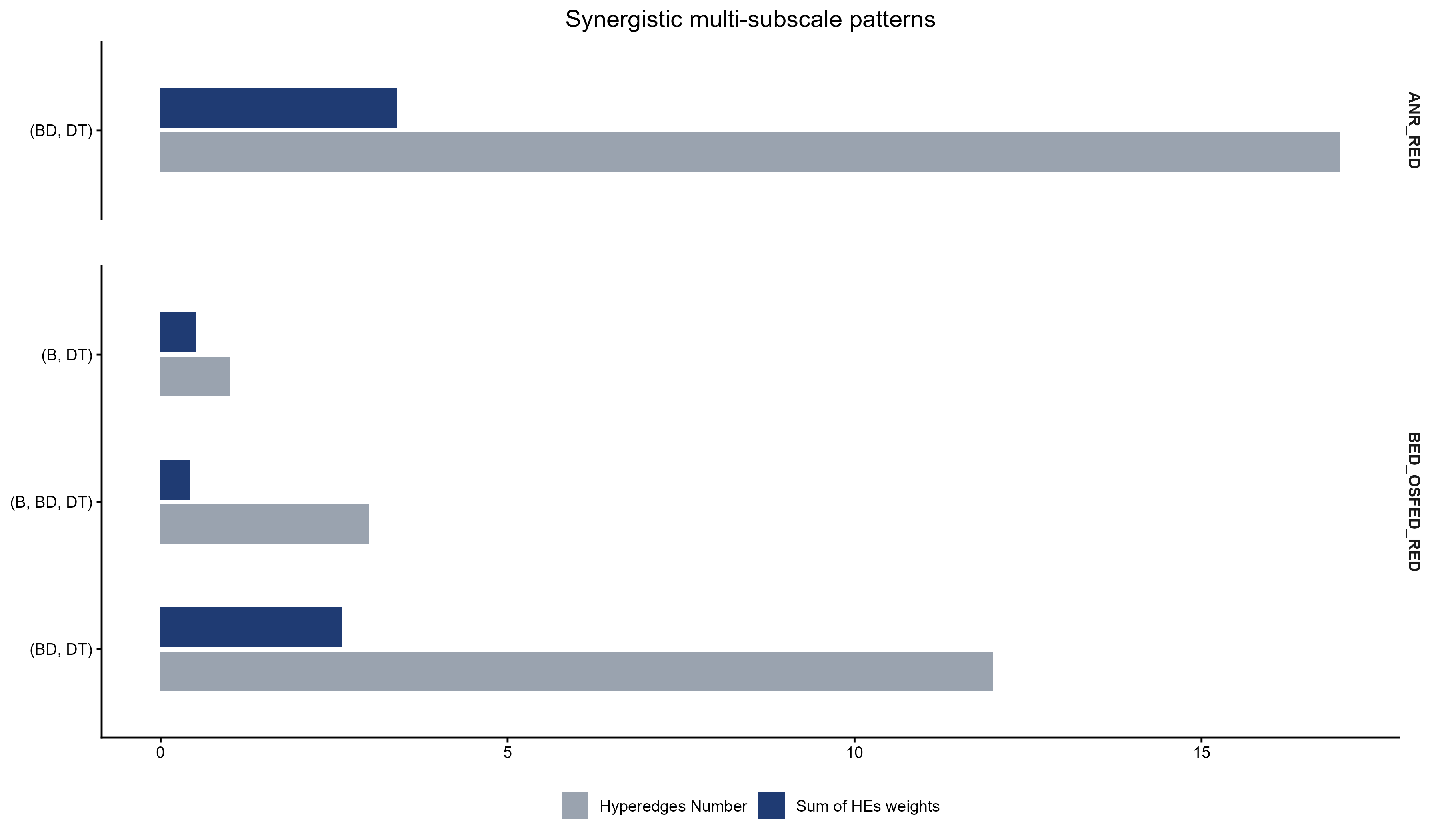} 
    \caption{\textbf{Redundant multi-subscale interaction patterns across diagnostic layers.} Two panels display the validated multi-subscale redundant configurations for ANR and BED/OSFED. For each pattern, the grey bar represents the number of hyperedges -- as \textit{Hyperedges Number} -- and the blue one represents the cumulative redundant weight -- as \textit{Sum of HEs weights}. Multi-subscale redundancy is limited and largely confined to combinations of eating-related dimensions, particularly Body Dissatisfaction and Drive for Thinness, indicating that redundancy primarily reflects overlapping symptom content rather than cross-domain integration.}
    \label{fig:red_pat_2panels}
\end{figure}

Later, we examined higher-order redundant configurations, focusing on \textit{multi-subscale patterns} (Figure~\ref{fig:red_pat_2panels}).

Redundant interactions exhibit a substantially more constrained structure than synergistic ones, both in diversity and diagnostic specificity. Redundancy is highly concentrated, involving a limited number of configurations dominated by closely related symptom dimensions. This indicates that overlapping information is primarily confined to tightly coupled constructs rather than distributed across multiple psychological domains.

Multi-subscale redundancy is rare. In \textit{ANBP} and \textit{BN}, no cross-subscale redundant patterns are observed, indicating that redundancy remains entirely within single domains -- predominantly body-image content. In \textit{ANR}, only one multi-subscale configuration emerges, namely the pair (BD, DT), reflecting strong informational overlap between body dissatisfaction and drive for thinness. In \textit{BED/OSFED}, a small number of cross-subscale redundant patterns appear, again centred on combinations of BD, DT, and B. These results reinforce the conclusion that redundancy arises from closely related eating-symptom dimensions, without extending to broader cross-domain integration.

\section*{Discussion}
We investigated eating-disorder psychopathology from a higher-order, systems-level perspective by modelling multivariate dependencies among EDI-3 items using an $\Omega$-Information framework. Validated interactions were organised into two multiplex hypergraphs -- one capturing synergistic effects and the other redundant ones -- across four diagnostic layers defined on a shared set of 91 items. This design enabled direct cross-diagnostic comparison within a common measurement space and allowed the separation of emergent (synergistic) from overlapping (redundant) informational structures.
We addressed three main questions: whether structurally influential items are stable or diagnosis-specific; how subscales differ in their higher-order organisation across diagnoses; and whether synergy provides diagnostic insights. Particular attention was given to the potential of higher-order patterns to differentiate EDs subtypes.

\subsection*{Synergy as a marker of emergent, integrative structure}
Synergy reflects emergent higher-order dependence: combinations of symptoms convey more information jointly than through their individual contributions. Clinically, this implies that certain constellations of eating-related, affective, interpersonal, and self-evaluative symptoms form integrated patterns that cannot be reduced to isolated endorsements. The multiplex hypergraph framework formalizes this idea by representing synergy as higher-order configurations rather than simple overlap.
Across diagnoses, synergistic organization shows both stability and specificity. Interoceptive Deficits (ID) consistently acts as a transdiagnostic backbone at both item and subscale levels, indicating that internal-state processing and affective awareness are central to higher-order ED structure. However, other dimensions vary in prominence across layers. ANBP shows a relatively circumscribed integration of interoceptive, developmental, and eating-related themes; ANR emphasizes restrictive control alongside interoception and maturity concerns; BED/OSFED displays a more heterogeneous and distributed configuration; and BN is characterized by strong interpersonal integration and more complex, cross-domain higher-order patterns.
At the item level, influential synergistic nodes repeatedly arise from ID, LSE, DT, and MF, suggesting a shared structural core. Yet the specific composition and integration of these items differ across diagnoses, indicating that synergy is neither fully invariant nor entirely diagnosis-specific.
Overall, synergy has a dual nature: it encodes a stable transdiagnostic foundation while also revealing diagnosis-sensitive reconfigurations in how symptom domains integrate. Differences in the order and complexity of synergistic interactions further suggest that higher-order groupings may serve as informative structural signatures of EDs subtypes. Compared to redundancy -- largely confined to overlapping eating-related content -- and dyadic networks, synergy provides added value by capturing emergent multivariate organization and the integrative architecture of psychopathology.

\subsection*{Redundancy as a marker of overlap and domain-specific coherence}
In contrast to synergy, redundant structure is highly concentrated, involves few configurations, and is dominated by a limited set of closely related subscales -- indicating overlap within specific symptom domains rather than cross-domain integration.
Across diagnostic layers, redundancy is consistently centred on eating and body-image dimensions, particularly Body Dissatisfaction (BD), with secondary contributions from Drive for Thinness (DT) and Bulimia (B). In ANBP and BN, redundant organization is almost entirely confined to BD -- with B still taking a role -- suggesting strong internal coherence among body-image evaluations. ANR shows broader redundancy shared between BD and DT, reflecting overlap between body dissatisfaction and weight-control concerns. BED/OSFED presents the widest distribution, with redundancy spanning BD, DT, and B, yet still restricted to closely related eating-symptom content.
Multi-subscale redundant patterns are rare. ANBP and BN show none, ANR exhibits only the BD–DT pair, and BED/OSFED presents a small number of configurations limited to eating-related scales. 
These findings reinforce the conclusion that redundancy captures informational duplication within tightly coupled constructs rather than higher-order integration across psychological domains -- acting as a marker of domain-specific coherence and interchangeability.

\subsection*{Clinical implications and further studies}
The present findings have several relevant clinical implications: by identifying higher-order, multi-symptom interaction patterns that go beyond traditional pairwise associations, this approach provides a more nuanced representation of eating disorder psychopathology. Clinically, this may support more targeted and personalized interventions, as treatment could focus not only on central symptoms but also on synergistic constellations of symptoms that mutually reinforce one another. For instance, interventions addressing co-occurring cognitive and affective components (e.g., overvaluation of weight/shape combined with emotion regulation difficulties) may prove more effective than strategies targeting isolated features. Moreover, the identification of diagnosis-specific versus transdiagnostic higher-order patterns may improve how clinicians understand each case and support more informed treatment planning across ED presentations.
Future research should aim to validate these findings in larger and more diverse clinical samples, as well as in longitudinal designs to assess the temporal stability and causal relevance of higher-order interactions. Integrating this framework with treatment outcome data could clarify whether specific synergistic configurations predict prognosis or response to particular therapeutic approaches. Additionally, extending the model to multimodal data (e.g., behavioral, neurobiological, and ecological momentary assessment data) may further enhance the ecological validity and clinical utility of higher-order network models. Connecting advanced psychometric models with clinical practice will be key to turning these insights into practical tools for assessment and treatment.

\section*{Methods}
\subsection*{Hyperedge estimation in multiplex layers}
The first challenge we addressed was the selection of hyperedges candidates due to the exploding number of possible combinations of multiplets of order $(3, 4, 5)$ that can be obtained out of $N = 91$ nodes -- the questionnaire's items. Only selected candidates were tested to check their synergistic, redundant or neutral nature. 
We followed two different strategies:
\begin{enumerate}
    \item \textbf{Network-Based Candidates from Dyadic Associations}: strong pairwise dependencies provide a structure that increases the likelihood of higher-order interactions, while avoiding the need to test all possible variable combinations. Candidate multiplets were derived from dyadic networks estimated using alternative correlation models, chosen to accommodate the distributional properties of ordinal psychometric data ~\citep{rhemtulla2012cat, johal2023net}: \textbf{nonparanormal correlation}~\citep{liu2009nonpara} and \textbf{polychoric correlation}~\citep{holgado2010poly, fox2016polycor, kiwanuka2022poly}. 
    
    For both correlation estimation strategies, network structure was inferred using the Extended Bayesian Information Criterion graphical lasso (\emph{EBICglasso}), yielding sparse partial correlation networks. 
    This dual estimation approach ensured robustness to violations of distributional assumptions while maintaining consistency across diagnostic groups.

    From each dyadic network -- signed, weighted, undirected and connected -- candidate multiplets were selected by leveraging its mesoscale organization.
    First, SpinGlass community detection yielded hard community memberships and an inter-community affinity matrix \(W\), defined as the mean absolute coupling between communities. This algorithm is well suited for weighted networks with signed edges and has been shown to perform reliably on graphs of moderate size, such as psychometric item networks(~\citep{yang2016spin, punzi2023network}) Candidate sets were seeded from maximal cliques (~\citep{gao2022cliques, contisciani2022cliques}) following a community-informed strategy (~\citep{contisciani2022cliques}). Cliques -- restricted to \([k_{\min}, k_{\max}]\), with large cliques decomposed -- were scored as
    \begin{equation}
    S(e) = \kappa(|e|)\sum_{i<j \in e} u_i^{\top} W u_j,
    \end{equation}
    which rewards sets spanning strongly interacting communities while penalizing large sizes via \(\kappa(|e|)=1/|e|!\). Top seeds were then greedily expanded by adding neighboring nodes that produced a sufficient increase in the structural score, thereby capturing near-clique structures. Finally, the resulting multiplets were grouped by order.

    \item \textbf{Subscale-Guided Combinatorial Candidates.} We exploited prior psychometric structure encoded in EDI-3 subscales to select hyperedges candidates. We considered intra-subscale combinations, which include items from the same subscale, and tested whether items designed to measure a common construct exhibit redundant or synergistic structure beyond pairwise associations. Then, we analysed inter-subscale combinations, which couple items from different subscales to probe cross-domain interactions, ensuring each multiplet contained at least one item from each subscale.
    Because inter-subscale combinations increase combinatorially with interaction order, all third-order candidates were retained, whereas higher-order combinations were randomly sampled within each subscale pair. This strategy balanced exhaustive coverage at the lowest non-trivial order with computational tractability at higher orders.
\end{enumerate}

\subsection*{\(\Omega\)-information hyperedge validation and selection.}
Given a multivariate system \(X=\{X_1,\dots,X_N\}\) (EDI-3 items), the global \(\Omega(X_1,\dots,X_N)\) summarizes the balance between redundancy and synergy but can mask localized higher-order effects. To identify statistically meaningful interactions, \(\Omega\)-information was therefore evaluated on candidate multiplets of order \(k\in\{3,4,5\}\).
For each multiplet, \(\Omega\) was computed and validated through a three-stage significance test, following the approach introduced by Marinazzo \emph{et al.} (~\citep{marinazzo2024info}): 
\emph{Stage 1} performed a column-wise permutation test: each variable was independently shuffled to preserve marginal distributions while destroying multivariate dependence, yielding a null distribution for \(\Omega\); two-tailed p-values were computed, adjusted for multiple comparisons (FDR via Benjamini--Hochberg~\citep{benjamini1995fdr}), and combined with an effect-size tolerance (e.g., treating \(|\Omega|<0.15\) as negligible) to retain only non-trivial significant multiplets. 
\emph{Stage 2} checked robustness via row-wise nonparametric bootstrap resampling and BCa confidence intervals (~\citep{diciccio1996bca, efron1994boot, efron2020bca}), discarding multiplets whose intervals overlap zero (unstable sign) or whose point estimates appear unreliable relative to the bootstrap distribution. 
\emph{In Stage 3}, a hierarchical check compared each \(k\)-multiplet confidence interval with those of all \((k-1)\)-submultiplets, removing sets whose intervals overlap, indicating no added higher-order effect beyond lower-order structure (~\citep{marinazzo2024info}).

This pipeline yielded a conservative set of statistically significant, stable, and genuinely higher-order redundant (\(\Omega>0\)) or synergistic (\(\Omega<0\)) multiplets suitable for hypergraph construction.

\subsection*{Diagnostic multiplex layers construction.}
The full candidate generation and \(\Omega\)-information validation workflow was applied independently to each diagnostic dataset to obtain diagnosis-specific sets of synergistic and redundant multiplets. For each group, dyadic networks were first estimated, candidate multiplets were produced via both network-based and subscale-guided strategies, and all candidates were evaluated through the inferential pipeline combining permutation-based null-model testing, BCa bootstrap confidence intervals, and cross-order confidence-interval comparisons. Multiplets that passed all filtering steps were retained as statistically validated higher-order interactions and interpreted as hyperedges characterizing the informational structure of the corresponding diagnosis. The resulting diagnostic-specific hyperedges were then organized into a multiplex representation, where each layer corresponds to a diagnosis and contains its validated hyperedges, enabling direct comparison of shared versus diagnosis-specific interaction patterns. Because the same pipeline is used across all diagnostic datasets, observed differences between layers can be attributed to genuine variation in informational organization rather than methodological artifacts. This procedure yielded two multiplex hypergraphs, one encoding synergistic hyperedges and one encoding redundant hyperedges.

\subsection*{Methodological validation via Synthetic Data}
We performed methodological validation using synthetic datasets with
\(\Omega\)-Information of known sign and magnitude, following the simulation framework
of Marinazzo \emph{et al.} ~\citep{marinazzo2024info}. This approach generates multivariate Gaussian systems by:
(i) constructing triplets \((X,Y,Z)\) whose \(3\times 3\) correlation matrices produce a
target \(\Omega\); (ii) assembling multiple independent triplets into a larger system via a
block-diagonal covariance matrix; and (iii) sampling and standardizing Gaussian data to
obtain correlation matrices for analysis. Control over \(\Omega\) is achieved through a
single-factor parameterization in which higher-order effects are tuned by varying a
residual (error) covariance between two observed variables, allowing precise modulation
of redundancy (\(\Omega>0\)) or synergy (\(\Omega<0\)) without assuming a specific causal
mechanism. Using nine triplets per dataset, four regimes were simulated: near-zero
interaction (\(\Omega\approx 0\), \(e_{\mathrm{cov}}=-0.15\)), redundant
(\(e_{\mathrm{cov}}=0.22\)), synergistic (\(e_{\mathrm{cov}}=-0.39\)), and mixed systems
combining all three regimes across triplets.

\subsection*{Quantifying items and subscales functional involvement}
After developing the pipeline to estimate the multiplex hypergraphs, we defined a measure to quantify the functional involvement of psychometric objects within each diagnostic layer. Starting from individual items, we computed the weighted degree of every node in both the synergy and redundancy multiplex hypergraphs. This measure captures the total strength of higher-order interactions incident to a node and provides an indicator of its informational role within a given diagnosis.

Formally, let $G^{\alpha}$ denote the hypergraph associated with layer $\alpha$, and let $H \in \{0,1\}^{N \times E}$ be its incidence matrix, where $H_{ie} = 1$ if node $i$ belongs to hyperedge $e$, and $0$ otherwise. Let $\mathbf{w} \in \mathbb{R}^{E}$ be the vector of hyperedge weights, with entry $w_e$ corresponding to hyperedge $e$. The weighted degree of node $i$ in layer $\alpha$ is defined as
\begin{equation}
k^{(w)}_{i} = \sum_{e=1}^{E} H_{ie} \, w_e,
\end{equation}
that is, the total weight of all hyperedges incident to node $i$.

To enable meaningful comparisons across layers, we normalized the weighted degree by the total absolute interaction strength within each layer. This normalization accounts for differences in overall magnitude of higher-order dependencies and yields a layer-specific indicator of the relative importance of each node. Since synergy and redundancy are analyzed in separate multiplexes, and all hyperedges within a layer share the same sign (either all positive or all negative), no additional sign correction was required.

To quantify the involvement of each EDI-3 \textit{subscale} in higher-order interactions, we defined a scale-level score derived from node weighted degrees. For a given diagnostic layer $\alpha$, let $\tilde{k}^{(w)}_{i}$ denote the normalized weighted degree of item $i$. The raw scale contribution was computed as the sum of weighted degrees of all items belonging to scale $s$.

To account for differences in the number of items across subscales, this quantity was normalized by the scale size $|s|$, yielding an average per-item contribution:
\begin{equation}
\bar{S}_s = \frac{1}{|s|} \sum_{i \in s} \tilde{k}^{(w)}_{i}.
\end{equation}

Finally, to enable comparison across scales within the same layer, the scale scores were rescaled by the total across all subscales:
\begin{equation}
v_s = \frac{\bar{S}_s}{\sum_r \bar{S}_r}.
\end{equation}

The resulting normalized scale score $v_s$, which we called Normalized Scale Weighted Degree (\textit{NSWD}), provides a layer-specific measure of the relative participation of each subscale in the higher-order synergistic or redundant structure, while controlling for unequal scale lengths.

To quantify the relevance of interaction patterns at the subscale level, we derived a ranking based on their involvement in the hypergraph structure. Specifically, patterns were ranked primarily according to the cumulative weight of all hyperedges in which the pattern appears, capturing the overall strength of its higher-order associations. In cases of equal cumulative weight, patterns were further ordered by the number of hyperedges in which they are present, reflecting their frequency of occurrence across higher-order interactions. This ranking procedure prioritizes patterns that are both strongly and consistently involved in the higher-order organization of the network.

\section*{Data Availability}
\label{data}
This work uses the Eating Disorder Inventory-3 (EDI-3), a proprietary psychometric instrument published by Psychological Assessment Resources (PAR, Inc.)~\citep{garner2004edi3}. In compliance with copyright and licensing restrictions, full item wording, scoring procedures, and verbatim content from the EDI-3 manual are not reproduced. Subscale names referenced throughout the work (e.g., Drive for Thinness, Bulimia, Body Dissatisfaction) are reported solely for scholarly description and interpretation. All rights to the EDI-3 are held by the original publisher~\citep{garner2004edi3}.

The clinical dataset analyzed in this study is the Eating Disorder Inventory-3, American clinical cases~\citep{brookings2020edi3}.
The dataset is publicly available from the Inter-university Consortium for Political and Social Research (ICPSR) openICPSR repository (DOI: \url{https://doi.org/10.3886/E109443V2}). Access is subject to ICPSR registration and terms of use.
Supplementary information regarding diagnostic labels were provided courtesy of Dr.\ David M.\ Garner~\citep{garner2004edi3}, author of the EDI-3 and co-supervisor of the dataset collection.

\newpage

\begin{table*}[t]
\centering
\caption{Top items ranked by normalized weighted degree in the synergistic layers across diagnoses.}
\label{tab:syn_items_all}
\renewcommand{\arraystretch}{1.15}
\footnotesize
\begin{tabular}{cccp{7.5cm}}
\hline
\textbf{Item} & \textbf{Norm. WD} & \textbf{Subscale} & \textbf{Item Descriptor} \\
\hline
\multicolumn{4}{l}{\textbf{ANBP\_SYN}} \\
\hline
8  & 0.0503 & ID  & Fear of intense emotions \\
10 & 0.0496 & LSE & Reduced self-worth \\
2  & 0.0398 & BD  & Stomach size dissatisfaction \\
6  & 0.0324 & MF  & Fear of aging \\
1  & 0.0318 & DT  & Anxiety about sweets/starches \\
3  & 0.0283 & MF  & Longing for childhood safety \\
35 & 0.0250 & MF  & Fears of adulthood \\
5  & 0.0243 & B   & Binge eating \\
40 & 0.0220 & ID  & Hunger/satiety confusion \\
44 & 0.0196 & ID  & Fear of emotional loss of control \\
\hline
\multicolumn{4}{l}{\textbf{ANR\_SYN}} \\
\hline
8  & 0.0808 & ID & Fear of intense emotions \\
1  & 0.0739 & DT & Anxiety about sweets/starches \\
6  & 0.0403 & MF & Fear of aging \\
2  & 0.0388 & BD & Stomach size dissatisfaction \\
7  & 0.0375 & DT & Dieting intention \\
3  & 0.0361 & MF & Longing for childhood safety \\
33 & 0.0270 & ID & Difficulty identifying emotions \\
40 & 0.0258 & ID & Hunger/satiety confusion \\
10 & 0.0226 & LSE & Reduced self-worth \\
47 & 0.0219 & BD & Discomfort after normal intake \\
\hline
\multicolumn{4}{l}{\textbf{BED\_OSFED\_SYN}} \\
\hline
8  & 0.0499 & ID & Fear of intense emotions \\
1  & 0.0435 & DT & Anxiety about sweets/starches \\
10 & 0.0348 & LSE & Reduced self-worth \\
44 & 0.0302 & ID & Fear of emotional loss of control \\
33 & 0.0291 & ID & Difficulty identifying emotions \\
77 & 0.0284 & ID & Intrusive thoughts / obsessions \\
40 & 0.0265 & ID & Hunger/satiety confusion \\
6  & 0.0405 & MF & Fear of aging \\
7  & 0.0227 & DT & Dieting intention \\
54 & 0.0221 & IA & Interpersonal distance / boundary discomfort \\
\hline
\multicolumn{4}{l}{\textbf{BN\_SYN}} \\
\hline
8  & 0.0474 & ID & Fear of intense emotions \\
10 & 0.0463 & LSE & Reduced self-worth \\
2  & 0.0325 & BD & Stomach size dissatisfaction \\
1  & 0.0311 & DT & Anxiety about sweets/starches \\
89 & 0.0309 & IA & Feeling loved / perceived support \\
6  & 0.0296 & MF & Fear of aging \\
17 & 0.0289 & IA & Interpersonal trust and awareness \\
69 & 0.0254 & II & Social ease in groups \\
18 & 0.0250 & PA & Loneliness / social isolation \\
54 & 0.0246 & IA & Interpersonal distance / boundary discomfort \\
\hline
\end{tabular}
\end{table*}

\begin{table*}[t]
\centering
\caption{Top items ranked by normalized weighted degree in the synergistic layers across diagnoses.}
\label{tab:red_items_all}
\renewcommand{\arraystretch}{1.15}
\footnotesize
\begin{tabular}{cccp{7.5cm}}
\hline
\textbf{Item} & \textbf{Norm. WD} & \textbf{Subscale} & \textbf{Item Descriptor} \\
\hline
\multicolumn{4}{l}{\textbf{ANBP\_RED}} \\
\hline
55 & 0.1516 & BD & Thigh-size satisfaction \\
62 & 0.1346 & BD & Hip-width satisfaction \\
45 & 0.1326 & BD & Hips size dissatisfaction \\
9  & 0.0786 & BD & Thighs size dissatisfaction \\
12 & 0.0758 & BD & Belly size satisfaction \\
19 & 0.07   & BD & Body shape satisfaction \\
59 & 0.0557 & BD & Buttocks size dissatisfaction \\
2 & 0.0496  & BD & Stomach size dissatisfaction \\
28 & 0.0424 & B  & Binge episodes \\
5  & 0.0413 & B  & Binge eating \\
\hline
\multicolumn{4}{l}{\textbf{ANR\_RED}} \\
\hline
32 & 0.1464 & DT & Preoccupation for thinness \\
49 & 0.1115 & DT & Fear of ongoing weight gain \\
16 & 0.1104 & DT & Fear of weight gain \\
9  & 0.1101 & BD & Thighs size dissatisfaction \\
7  & 0.0822 & DT & Dieting intention \\
2  & 0.0777 & BD & Stomach size dissatisfaction \\
45 & 0.072  & BD & Hips size dissatisfaction \\
62 & 0.0549 & BD & Hip-width satisfaction \\
55 & 0.0522 & BD & Thigh-size satisfaction \\
59 & 0.0418 & BD & Buttocks size dissatisfaction \\
\hline
\multicolumn{4}{l}{\textbf{BED\_OSFED\_RED}} \\
\hline
5 & 0.0828 & B & Binge eating \\
62 & 0.069 & BD & Hip-width satisfaction \\
46 & 0.0646 & B & Secretive overeating \\
4  & 0.0629 & B & Urge to eat when distressed \\
38 & 0.0572 & B & Preoccupation with binge eating \\
45 & 0.0559 & BD & Hips size dissatisfaction \\
9 & 0.0548 & BD & Thighs size dissatisfaction \\
2 & 0.053 & BD & Stomach size dissatisfaction \\
28 & 0.052  & B & Binge episodes \\
12 & 0.0479 & BD & Belly size satisfaction \\
\hline
\multicolumn{4}{l}{\textbf{BN\_RED}} \\
\hline
55 & 0.1809 & BD & Thigh-size satisfaction \\
62 & 0.1589 & BD & Hip-width satisfaction \\
45 & 0.1209 & BD & Hips size dissatisfaction \\
9  & 0.1172 & BD & Thighs size dissatisfaction \\
19 & 0.0807 & BD & Body shape satisfaction \\
12 & 0.0766 & BD & Belly size satisfaction \\
59 & 0.0744 & BD & Buttocks size dissatisfaction \\
2  & 0.0515 & BD & Stomach size dissatisfaction \\
31 & 0.0376 & BD & Buttocks shape satisfaction \\
21 & 0.0201 & BD & Difficulty identifying emotions \\
\hline
\end{tabular}
\end{table*}

\newpage

\begin{thebibliography}{99}


\bibitem[Garner(2004)]{garner2004edi3}
Garner, D. M. (2004).
\textit{Eating Disorder Inventory--3: Professional manual}.
Psychological Assessment Resources.

\bibitem[Brookings(2020)]{brookings2020edi3}
Brookings, Jeffrey. (2020)
Eating Disorder Inventory-3, American clinical cases. Ann Arbor, MI: \textit{Inter-university Consortium for Political and Social Research} [distributor], 2020-06-04. https://doi.org/10.3886/E109443V2

\bibitem[American Psychiatric Association(2022)]{dsm5tr}
American Psychiatric Association. (2022).
\textit{Diagnostic and statistical manual of mental disorders}
(5th ed., text rev.; DSM-5-TR).
American Psychiatric Publishing.

\bibitem[Punzi et al.(2023)]{punzi2023network}
Punzi, C., et al. (2023).
Network-based validation of the EDI-3 for the assessment of eating disorders.
\textit{Scientific Reports}, 13, 1578.
\texttt{https://doi.org/10.1038/s41598-023-28641-6}

\bibitem[Feng et al.(2023)]{feng2023eds}
Feng, B., Harms, J., Chen, E., Gao, P., Xu, P., \& He, Y. (2023).
Current discoveries and future implications of eating disorders.
\textit{International Journal of Environmental Research and Public Health},
20(14), 6325.
\texttt{https://doi.org/10.3390/ijerph20146325}

\bibitem[Levinson et al.(2024)]{levinson2024guide}
Levinson, C. A., Osborn, K., Hooper, K., Vanzhula, I., \& Ralph-Nearman, C. (2024).
Evidence-based assessments for transdiagnostic eating disorder symptoms:
Guidelines for current use and future directions.
\textit{Assessment}, 31(1), 145--167.
\texttt{https://doi.org/10.1177/10731911231201150}

\bibitem[Barakat et al.(2023)]{barakat2023risk}
Barakat, S., McLean, S. A., Bryant, E., et al. (2023).
Risk factors for eating disorders: Findings from a rapid review.
\textit{Journal of Eating Disorders}, 11, 8.
\texttt{https://doi.org/10.1186/s40337-022-00717-4}

\bibitem[Clausen et al.(2011)]{clausen2011val}
Clausen, L., Rosenvinge, J. H., Friborg, O., \& Rokkedal, K. (2011).
Validation of the Eating Disorder Inventory--3 in a clinical sample.
\textit{International Journal of Eating Disorders}, 44(7), 642--646.

\bibitem[Bhattacharya et al.(2020)]{bhatta2020feed}
Bhattacharya, A., DeFilipp, L., \& Timko, C. A. (2020).
Chapter 24 --- Feeding and eating disorders.
In R. Lanzenberger et al. (Eds.),
\textit{Handbook of Clinical Neurology} (Vol. 175, pp. 387--403).
Elsevier.

\bibitem[Ortiz et al.(2026)]{ortiz2026eds}
Ortiz, A. M. L., Peters, A. S., Webber, K. T., Butler, R. M., Fitterman-Harris, H. F., \& Levinson, C. A. (2026).
Eating disorder symptoms and corresponding evidence-based treatments: A narrative review.
\textit{Journal of Eating Disorders}, 14(1), 45.
\texttt{https://doi.org/10.1186/s40337-025-01485-7}

\bibitem[Eddy et al.(2008)]{eddy2008crossover}
Eddy, K. T., Dorer, D. J., Franko, D. L., Tahilani, K., Thompson-Brenner, H., \& Herzog, D. B. (2008).
Diagnostic crossover in anorexia nervosa and bulimia nervosa: Implications for DSM-V.
\textit{American Journal of Psychiatry}.

\bibitem[Fairburn et al.(2003)]{fairburn2003transdiagnostic}
Fairburn, C. G., Cooper, Z., \& Shafran, R. (2003).
Cognitive behaviour therapy for eating disorders: A transdiagnostic theory and treatment.
\textit{Behaviour Research and Therapy}, 41(5), 509--528.


\bibitem[Borsboom(2017)]{borsboom2017network}
Borsboom, D. (2017).
A network theory of mental disorders.
\textit{World Psychiatry}, 16(1), 5--13.
\texttt{https://doi.org/10.1002/wps.20375}

\bibitem[Friedman et al.(2008)]{friedman2008sparse}
Friedman, J., Hastie, T., \& Tibshirani, R. (2008).
Sparse inverse covariance estimation with the graphical lasso.
\textit{Biostatistics}, 9(3), 432--441.
\texttt{https://doi.org/10.1093/biostatistics/kxm045}

\bibitem[Epskamp et al.(2018)]{epskamp2018estim}
Epskamp, S., Borsboom, D., \& Fried, E. I. (2018).
Estimating psychological networks and their accuracy: A tutorial paper.
\textit{Behavior Research Methods}, 50(1), 195--212.
\texttt{https://doi.org/10.3758/s13428-017-0862-1}

\bibitem[Borsboom and Cramer(2013)]{borsboom2013net}
Borsboom, D., \& Cramer, A. O. J. (2013).
Network analysis: An integrative approach to the structure of psychopathology.
\textit{Annual Review of Clinical Psychology}, 9, 91--121.
\texttt{https://doi.org/10.1146/annurev-clinpsy-050212-185608}

\bibitem[McNally(2016)]{mcnally2016brat}
McNally, R. J. (2016).
Can network analysis transform psychopathology?
\textit{Behaviour Research and Therapy}, 86, 95--104.
\texttt{https://doi.org/10.1016/j.brat.2016.06.006}

\bibitem[Benson et al.(2018)]{benson2018simplicial}
Benson, A. R., Abebe, R., Schaub, M. T., Jadbabaie, A., \& Kleinberg, J. (2018).
Simplicial closure and higher-order link prediction.
\textit{Proceedings of the National Academy of Sciences}, 115(48), E11221--E11230.
\texttt{https://doi.org/10.1073/pnas.1800683115}

\bibitem[Battiston et al.(2021)]{battiston2021natphys}
Battiston, F., et al. (2021).
The physics of higher-order interactions in complex systems.
\textit{Nature Physics}, 17(10), 1093--1098.
\texttt{https://doi.org/10.1038/s41567-021-01371-4}

\bibitem[Punzi et al.(2022)]{punzi2022review}
Punzi, C., Petti, M., \& Tieri, P. (2022).
Network-based methods for psychometric data of eating disorders: A systematic review.
\textit{PLoS One}.
\texttt{https://doi.org/10.1371/journal.pone.0276341}

\bibitem[Marinazzo et al.(2024)]{marinazzo2024info}
Marinazzo, D., Van Roozendaal, J., Rosas, F. E., Stella, M., Comolatti, R.,
Colenbier, N., Stramaglia, S., \& Rosseel, Y. (2024).
An information-theoretic approach to build hypergraphs in psychometrics.
\textit{Behavior Research Methods}, 56(1).
\texttt{https://doi.org/10.3758/s13428-023-02262-4}

\bibitem[Choo et al.(2025)]{choo2025hyper}
Choo, H., Bu, F., Hwang, H., Yoon, Y., \& Shin, K. (2025).
HyperSearch: Prediction of new hyperedges through unconstrained yet efficient search.
\textit{ICDM}.
\texttt{https://doi.org/10.48550/arXiv.2510.17153}

\bibitem[Christensen et al.(2020)]{christensen2020uva}
Christensen, A. P., Garrido, L. E., \& Golino, H. (2020).
Unique variable analysis: A network psychometrics method to detect local dependence.
\textit{PsyArXiv}.
\texttt{https://doi.org/10.31234/osf.io/4kra2}


\bibitem[Lotito et al.(2024)]{lotito2024multiplex}
Lotito, Q. F., Montresor, A., \& Battiston, F. (2024).
Multiplex measures for higher-order networks.
\textit{Applied Network Science}, 9, 55.
\texttt{https://doi.org/10.1007/s41109-024-00665-9}


\bibitem[Rosas et al.(2019)]{rosas2019oinformation}
Rosas, F. E., Mediano, P. A. M., Gastpar, M., \& Jensen, H. J. (2019).
Quantifying high-order interdependencies via multivariate information theory.
\textit{Physical Review E}, 100(3), 032305.
\texttt{https://doi.org/10.1103/PhysRevE.100.032305}


\bibitem[Rhemtulla et al.(2012)]{rhemtulla2012cat}
Rhemtulla, M., Brosseau-Liard, P. E., \& Savalei, V. (2012).
When can categorical variables be treated as continuous?
\textit{Psychological Methods}, 17(3), 354--373.
\texttt{https://doi.org/10.1037/a0029315}

\bibitem[Johal and Rhemtulla(2023)]{johal2023net}
Johal, S. K., \& Rhemtulla, M. (2023).
Comparing estimation methods for psychometric networks with ordinal data.
\textit{Psychological Methods}, 28(6), 1251--1272.
\texttt{https://doi.org/10.1037/met0000449}

\bibitem[Holgado-Tello et al.(2010)]{holgado2010poly}
Holgado-Tello, F. P., Chacón-Moscoso, S., Barbero-García, I., \& Vila-Abad, E. (2010).
Polychoric versus Pearson correlations in exploratory and confirmatory factor analysis.
\textit{Quality \& Quantity}, 44, 153--166.
\texttt{https://doi.org/10.1007/s11135-008-9190-y}

\bibitem[Kiwanuka et al.(2022)]{kiwanuka2022poly}
Kiwanuka, F., Kopra, J., Sak-Dankosky, J., Nanyonga, R. C., \& Kvist, T. (2022).
Polychoric correlation with ordinal data in nursing research.
\textit{Nursing Research}, 71(6), 460--469.
\texttt{https://doi.org/10.1097/NNR.0000000000000600}

\bibitem[Liu et al.(2009)]{liu2009nonpara}
Liu, H., Lafferty, J., \& Wasserman, L. (2009).
The nonparanormal: Semiparametric estimation of high-dimensional undirected graphs.
\textit{Journal of Machine Learning Research}, 10, 2295--2328.

\bibitem[Fox(2016)]{fox2016polycor}
Fox, J. (2016).
\textit{polycor}: Polychoric and polyserial correlations.
R package documentation.


\bibitem[Gao et al.(2022)]{gao2022cliques}
Gao, X., Zhou, F., Xu, K., Tian, X., \& Chen, Y. (2022).
A parallel algorithm for maximal cliques enumeration to improve hypergraph construction.
\textit{Journal of Computational Science}, 65, 101905.
\texttt{https://doi.org/10.1016/j.jocs.2022.101905}

\bibitem[Contisciani et al.(2022)]{contisciani2022cliques}
Contisciani, M., Battiston, F., \& De Bacco, C. (2022).
Inference of hyperedges and overlapping communities in hypergraphs.
\textit{Nature Communications}, 13, 7229.
\texttt{https://doi.org/10.1038/s41467-022-34714-7}

\bibitem[Yang et al.(2016)]{yang2016spin}
Yang, Z., Algesheimer, R., \& Tessone, C. J. (2016).
A comparative analysis of community detection algorithms on artificial networks.
\textit{Scientific Reports}, 6, 30750.
\texttt{https://doi.org/10.1038/srep30750}


\bibitem[Benjamini and Hochberg(1995)]{benjamini1995fdr}
Benjamini, Y., \& Hochberg, Y. (1995).
Controlling the false discovery rate: A practical and powerful approach to multiple testing.
\textit{Journal of the Royal Statistical Society: Series B}, 57, 289--300.

\bibitem[Efron and Tibshirani(1994)]{efron1994boot}
Efron, B., \& Tibshirani, R. J. (1994).
\textit{An introduction to the bootstrap}.
Chapman \& Hall/CRC.

\bibitem[DiCiccio and Efron(1996)]{diciccio1996bca}
DiCiccio, T. J., \& Efron, B. (1996).
Bootstrap confidence intervals.
\textit{Statistical Science}, 11(3), 189--228.

\bibitem[Davison and Hinkley(1997)]{davison1997boot}
Davison, A. C., \& Hinkley, D. V. (1997).
\textit{Bootstrap methods and their application}.
Cambridge University Press.

\bibitem[Efron and Narasimhan(2020)]{efron2020bca}
Efron, B., \& Narasimhan, B. (2020).
The automatic construction of bootstrap confidence intervals.
\textit{Journal of Computational and Graphical Statistics}, 29(3), 608--619.
\texttt{https://doi.org/10.1080/10618600.2020.1714633}

\end{thebibliography}
\end{document}